\documentclass[aps,prl,reprint,superscriptaddress,longbibliography]{revtex4-1}

\usepackage{bm,amsmath,amsfonts,amssymb,braket}
\usepackage{latexsym,dsfont,array,mathrsfs,textcase}
\usepackage{graphicx,xcolor}
\usepackage{comment,verbatim}
\usepackage{float}
\usepackage{times}
\usepackage{xspace}
\usepackage{stmaryrd}
\usepackage{multirow}
\usepackage{mathtools}
\usepackage{booktabs}
\usepackage[colorlinks,linkcolor=blue,citecolor=blue,urlcolor=blue]{hyperref}

\newcommand{\SM}{Supplemental Material~\cite{SM}}

\newcommand{\ii}{\mathrm{i}}
\newcommand{\e}{\mathrm{e}}

\begin{document}

\title{Lyapunov Exponents as Duality-Invariant Signatures of Critical States}

\author{Tong Liu}
\thanks{t6tong@njupt.edu.cn}
\affiliation{Department of Applied Physics, School of Science, Nanjing University of Posts and Telecommunications, Nanjing 210003, China}

\author{Gao Xianlong}
\thanks{gaoxl@zjnu.edu.cn}
\affiliation{Department of Physics, Zhejiang Normal University, Jinhua 321004, China}

\date{\today}

\begin{abstract}
Critical eigenstates are usually identified through wave-function geometry in a chosen basis, such as participation ratios, multifractal spectra, or finite-size scaling.  Here we formulate criticality instead as a dual-space Lyapunov property.  We prove a Fourier exclusion principle: exponential localization in one representation is incompatible with exponential localization in its Fourier-dual representation.  This turns the Liu--Xia condition, \(\gamma_x(E)=\gamma_m(E)=0\), from a phenomenological criterion into a rigorous length-scale statement: a critical state is characterized by the simultaneous absence of exponential confinement in real and momentum space.  The criterion is invariant under bounded local gauge transformations of the transfer matrix and remains compatible with conventional single-space multifractal diagnostics.  More importantly, it is exactly predictive.  In analytically tractable quasiperiodic models, the same condition yields closed-form critical lines, an exact finite critical region with an additional critical branch, and a complex critical surface in a non-Hermitian non-self-dual spectrum.  Thus the Liu--Xia condition provides not only a diagnostic of critical states, but an exact solvability principle for locating critical sets across distinct microscopic structures.
\end{abstract}

\maketitle

\textit{Introduction.---}
Invariants play an organizing role across physics because they reveal the part of a physical state that survives changes of description.  Gauge invariance tells us which quantities are observable rather than artifacts of a chosen phase convention~\cite{Berry1984}.  Topological invariants distinguish phases whose essential structure is invisible to local order parameters but robust against continuous deformations~\cite{TKNN1982}.  Scale-invariant quantities identify universal critical behavior by removing dependence on microscopic length scales~\cite{Wilson1975}.  In all these cases, the power of an invariant lies in its ability to separate intrinsic physics from representation-dependent detail. 

Localization transitions pose this question in a particularly sharp form~\cite{Anderson1958,IPR,Li1,Abrahams1979,MacKinnon1983,Kramer1993}.  A localized state is controlled by a finite exponential length, while an extended state fills a macroscopic part of the system.  A critical state lies between these two limits: it is neither confined to a finite localization volume nor uniformly spread over the entire lattice.  Its wave function instead carries nontrivial structure across scales.  Standard diagnostics such as participation ratios and multifractal dimensions describe this structure very well~\cite{Hiramoto1989,quasi2,EversMirlin2008,You}, but they are basis dependent.  A quantity that appears physically meaningful in one basis may acquire a completely different interpretation after a change of basis.

The physical issue is therefore not only whether a state looks critical in one representation, but whether criticality survives the comparison between two conjugate representations.  Lyapunov exponents provide the appropriate language because they measure the exponential rate of spatial growth or decay~\cite{IPR,Li1}.  A positive Lyapunov exponent means a finite localization length; a vanishing one means that exponential confinement is absent.  The Liu--Xia criterion uses this fact to characterize criticality by the simultaneous absence of exponential localization in real and momentum space~\cite{Liu-Critical}.  In this form, criticality is no longer a single-space geometric label, but a dual-space statement about the disappearance of characteristic exponential length scales.

This work develops this idea in three directions.  First, we prove a Fourier exclusion principle: a wave function exponentially localized in one representation cannot be exponentially localized in its Fourier-dual representation.  This establishes that the Liu--Xia condition is not a conjectural assumption, but a rigorous consequence.  Second, we clarify the relation to familiar critical signatures~\cite{Jiang2,Hofstadter1976,AubryAndre1980,Duncan,Ostlund1983,Avila,gg0,RG2}.  Multifractal scaling and divergent correlation lengths remain essential single-space probes~\cite{zhou,twochain}, but the dual-space Lyapunov condition isolates the invariant core that they share.  Third, we show that the condition is not merely diagnostic but exactly predictive.  For several analytically tractable models, the simultaneous conditions on the real- and dual-space Lyapunov exponents yield closed-form critical manifolds: energy-dependent critical lines in generalized self-dual Aubry--Andr\'e systems, a finite critical region together with a self-dual branch in class-dual decorated chains, and a complex critical surface in a non-Hermitian model without Aubry self-duality.  Thus the Liu--Xia condition does more than identify critical states after the fact; it provides an exact solvability principle for locating criticality across models with distinct microscopic structures~\cite{Longhi1,Zhang}.

\textit{Dual-space exclusion and the Liu--Xia condition.---}
\begin{figure*}[t]
\centering
\includegraphics[width=0.98\textwidth]{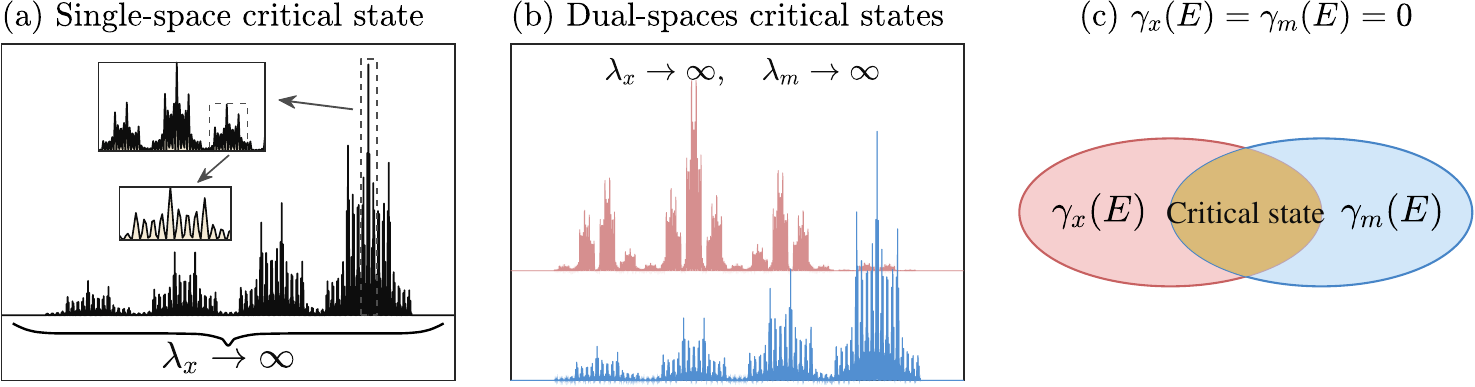}
\caption{
Schematic characterization of critical states.
(a) In a single representation, usually real space, criticality is associated with multifractal or self-similar wave-function structure, and a divergent real-space correlation length, \(\lambda_x\to\infty\).
(b) Dual-space criticality requires the simultaneous absence of a finite characteristic length in real and dual spaces, \(\lambda_x\to\infty\) and \(\lambda_m\to\infty\).
(c) In Lyapunov language, this is equivalent to the simultaneous vanishing of the corresponding inverse localization lengths, \(\gamma_x(E)=\gamma_m(E)=0\).
}
\label{fig:critical_state_schematic}
\end{figure*}

Consider a normalized state on a ring of length \(L\) and its discrete Fourier transform,
\begin{equation}
\phi_m=
\frac{1}{\sqrt L}
\sum_{n=1}^{L}\psi_n\e^{-\ii k_m n},
\qquad
k_m=\frac{2\pi m}{L} .
\label{eq:dft_main}
\end{equation}
The exclusion principle follows directly from this relation.  Suppose that \(\psi_n\) is exponentially localized around \(n_0\),
\[
|\psi_n|\sim \exp(-|n-n_0|/\xi),
\]
with finite localization length \(\xi\).  The Fourier sum then receives appreciable weight only from an \(O(\xi)\) neighborhood of \(n_0\), and therefore
\begin{equation}
|\phi_m|
\lesssim
\frac{1}{\sqrt L}
\sum_{|n-n_0|\lesssim \xi}|\psi_n|
=O(L^{-1/2}).
\label{eq:phi_scaling_main}
\end{equation}
Thus a state confined to only \(O(1)\) real-space sites necessarily spreads over \(O(L)\) components in its Fourier-dual momentum representation, and therefore cannot remain exponentially localized there.  By interchanging the two representations, the converse follows identically.  Exponential localization in one member of a Fourier-dual pair therefore excludes exponential localization in the other.

This dual-space exclusion admits a rigorous Lyapunov formulation through the standard transfer-matrix approach.  Unlike inverse participation ratios or information entropies, which depend on the chosen representation, Lyapunov exponents are invariant under bounded local gauge transformations of the transfer matrix and thus provide a robust diagnostic of exponential localization.  Let \(\gamma_x(E)\) and \(\gamma_m(E)\) denote the Lyapunov exponents of the real- and momentum-space transfer problems, respectively.  A positive exponent gives the inverse localization length in the corresponding representation.  Because exponential localization cannot occur simultaneously in a Fourier-dual pair, the critical case is singled out by the simultaneous absence of exponential confinement in both representations,
\begin{equation}
\gamma_x(E)=\gamma_m(E)=0 .
\label{eq:main-critical-lyapunov}
\end{equation}
This is the Liu--Xia condition.  It is not a phenomenological conjecture, but a dual-space length-scale criterion: both dual descriptions have zero exponential growth rate.  The rigorous proof of gauge invariance, together with the Fourier-exclusion argument, is given in Secs.~I and II of the \SM.

\paragraph{Compatibility with single-space critical signatures.---}
Critical states are often recognized in one representation through multifractal scaling or a divergent correlation length~\cite{Kohmoto1987,Last1994}.  These diagnostics are fully compatible with the Liu--Xia condition, but they play a different role.  They describe how the wave function distributes its weight inside a chosen basis, whereas the Lyapunov condition asks whether either basis contains a finite exponential localization length.

For example, multifractality~\cite{quasi4,quasi1} is encoded in the algebraic finite-size scaling
\begin{equation}
P_q^{(x)}(L)=\sum_n |\psi_n|^{2q}\sim L^{-\tau_q},
\end{equation}
and, in particular,
\begin{equation}
P_2^{(x)}(L)\sim L^{-D_2},
\qquad 0<D_2<1 .
\end{equation}
Such scaling means that the state is not confined to an \(O(1)\) localization volume, but it also does not behave as a featureless extended state.  The exponent \(D_2\) gives useful geometric information in that representation, while the vanishing Lyapunov exponent states the more basic fact that no exponential length cuts off the state.

The correlation-length formulation gives the same physical picture.  Since the localization length is the inverse Lyapunov exponent, the Liu--Xia condition requires the exponential localization lengths in both dual spaces to diverge.  As summarized in Fig.~\ref{fig:critical_state_schematic}, this turns the usual single-space picture of criticality into a dual-space statement: both conjugate representations must lack a finite exponential length scale.  Participation ratios and multifractal exponents then describe the internal geometry of this critical state in a chosen basis.

\paragraph{Generalized self-dual quasiperiodic model.---}
\begin{figure*}[t]
\centering
\includegraphics[width=0.98\textwidth]{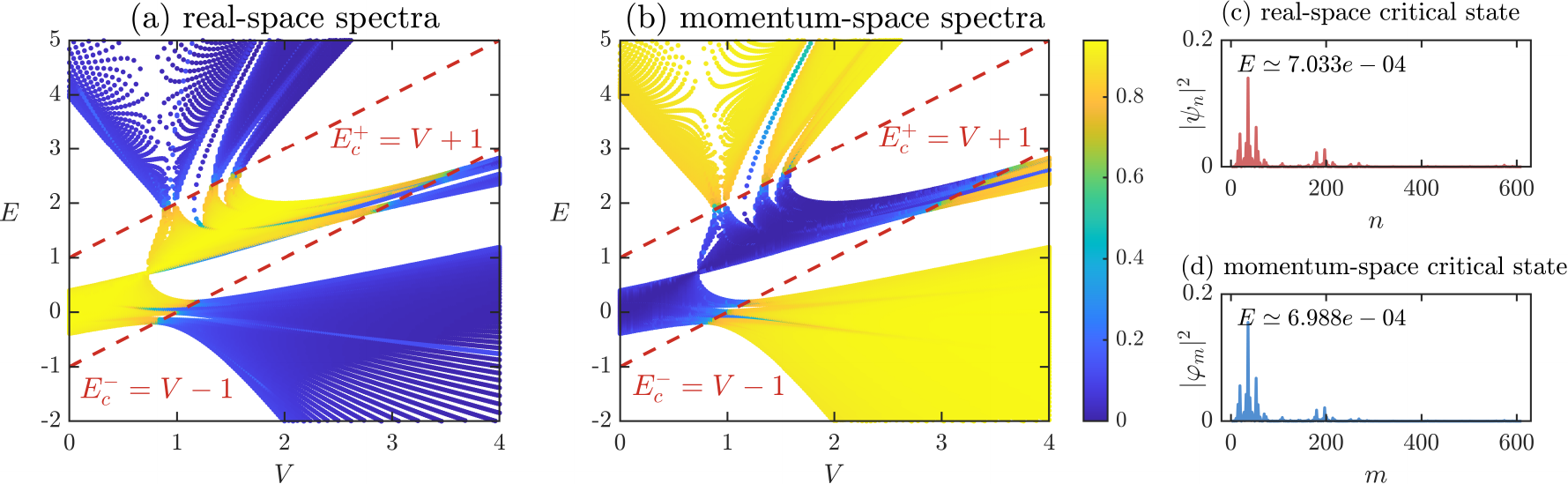}
\caption{
Dual-space spectra and the Liu--Xia critical set in a generalized self-dual model.
(a) Real-space and (b) momentum-space spectra as functions of \(V\), colored by the corresponding fractal dimension \(D_2\).
The calculation uses \(L=610\), \(\alpha=377/610\), and \(\mu=1\).
The dashed red lines, \(E_c^{-}=V-1\) and \(E_c^{+}=V+1\), mark the analytic boundaries predicted by the dual-space Lyapunov criterion.
(c),(d) Probability profiles of representative critical eigenstates at \(V=1\): \(\psi_n\) in real space and its Fourier-dual amplitude \(\phi_m\) in momentum space, with \(E\simeq 7.1\times10^{-4}\).  The numerical analysis is restricted to the purely real spectrum.
}
\label{fig:selfdual}
\end{figure*}

To show that the Liu--Xia condition not only characterizes critical states but also has exact predictive power, we first consider an analytically tractable quasiperiodic eigenvalue problem,
\begin{align}
&\psi_{n+1}+\psi_{n-1}
+2V\cos(2\pi\alpha n+\theta)\psi_n \\
&= E\psi_n
\nonumber
+2\mu E\cos(2\pi\alpha n+\theta)\psi_n ,
\end{align}
where \(V\) is the potential strength, \(\alpha\) is an irrational number that can be approximated by Fibonacci rational approximants, the phase is set to \(\theta=0\), \(\psi_n\) is the real-space wave-function amplitude at the \(n\)th lattice site, and \(\mu\) is an additional tunable parameter. 

It reduces exactly to an Aubry--Andr\'e equation with the energy-dependent effective potential~\cite{AubryAndre1980,Avila,gg0,RG2}
\(V_{\mathrm{eff}}(E)=V-\mu E\).
Using the Thouless formula~\cite{IPR,Li1}, the real- and momentum-space Lyapunov exponents are
\begin{align}
\gamma_x(E)&=\max\{0,\ln|V-\mu E|\}, \\
\gamma_m(E)&=\max\{0,-\ln|V-\mu E|\}.
\end{align}
These expressions explicitly realize the exclusion principle: the two Lyapunov exponents cannot be positive at the same energy. Imposing the Liu--Xia condition gives
\begin{equation}
E_c^{\pm}=\frac{V\pm1}{\mu},
\end{equation}
which predicts the energy-dependent critical lines before any wave-function scaling analysis is performed.  Compared with the standard Aubry--Andr\'e model, where self-duality fixes a single transition point~\cite{Roati2008}, the present model promotes duality into two spectrum-dependent critical branches.  Figure~\ref{fig:selfdual} confirms this prediction in both spectra and shows a representative eigenstate in the dual amplitudes \(\psi_n\) and \(\phi_m\). A rigorous derivation is given in Sec.~III of the \SM.

\begin{figure*}[t]
\centering
\includegraphics[width=0.98\textwidth]{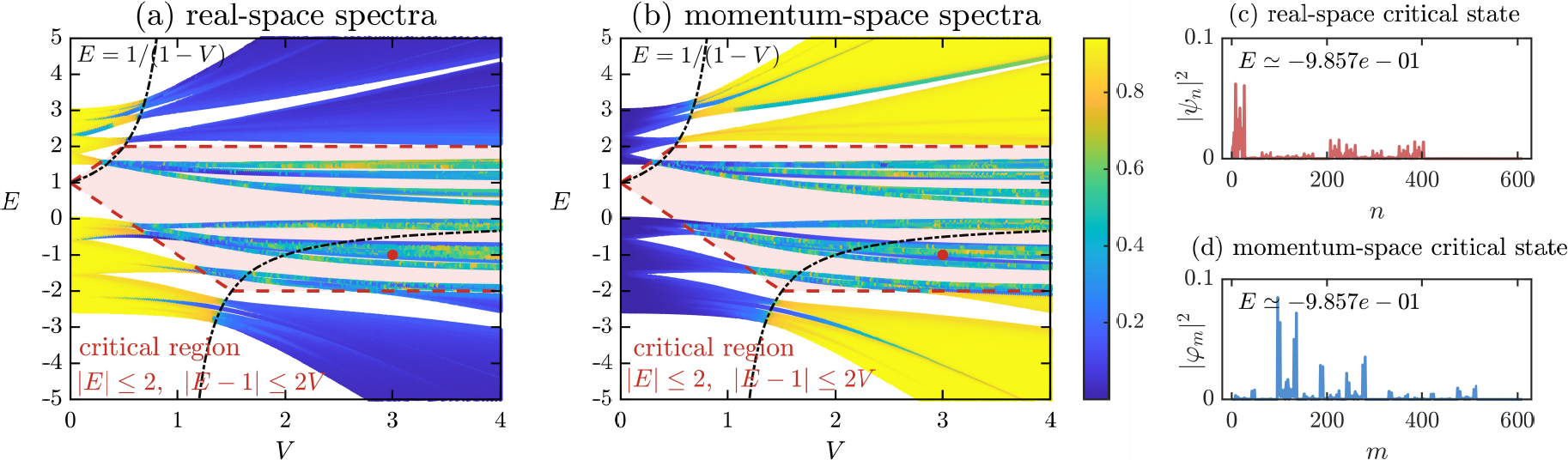}
\caption{
Liu--Xia critical set in a class-dual decorated-chain model.
(a),(b) Real- and momentum-space spectra as functions of \(V\), colored by the corresponding fractal dimension \(D_2\).
The calculations use \(L=610\), \(\alpha=377/610\), \(J=t=g=1\), and \(\theta=0\).
The shaded red region denotes the exact Liu--Xia critical set predicted from the criterion,
\(|E|\leq 2\) and \(|E-1|\leq 2V\).
The black dashed curve,
\(E=1/(1-V)\),
is the exact critical branch selected by the same criterion, consistent with the class-dual self-dual line. 
The red marker indicates a representative critical state ($V=3,~E=-1$), whose real- and momentum-space probability profiles are shown in (c),(d) through the amplitudes \(\psi_n\) and \(\phi_m\).  The numerical analysis is restricted to the purely real spectrum.
}
\label{fig:dual_sector}
\end{figure*}
\paragraph{Class-dual decorated-chain model.---}
We next consider a decorated quasiperiodic chain with one propagating backbone orbital \(A_n\) and a side-coupled symmetric dimer \((B_n,C_n)\) in each unit cell,
\begin{align}
H
=
\sum_n \Big[
&J(a_{n+1}^{\dagger}a_n+\mathrm{H.c.})
+g(a_n^{\dagger}b_n+a_n^{\dagger}c_n+\mathrm{H.c.})
\nonumber\\
&+t(b_n^{\dagger}c_n+\mathrm{H.c.})
+V_n(b_n^{\dagger}b_n+c_n^{\dagger}c_n)
\Big],
\label{eq:MC_H_main}
\end{align}
where \(V_n=2V\cos(2\pi\alpha n+ \theta)\).  The quasiperiodic modulation is applied only to the side dimer.  However, because a particle can virtually hop from the backbone to the side sites and back, eliminating the side modes transfers this modulation to the backbone as an energy-dependent quasiperiodic self-energy.  The backbone amplitudes therefore obey the exact real-space transport equation
\begin{equation}
a_{n+1}+a_{n-1}
=
\left[
\frac{E}{J}
-
\frac{2g^2/J}{E-t-2V\cos(2\pi\alpha n+\theta)}
\right]a_n .
\label{eq:MC_real_main}
\end{equation}
The Fourier-transformed nontrivial sector obeys
\begin{equation}
p_{k+1}+p_{k-1}
=
\left[
\frac{E-t}{V}
-
\frac{2g^2/V}{E-2J\cos(2\pi\alpha k+\theta)}
\right]p_k .
\label{eq:MC_dual_main}
\end{equation}
The model is therefore not self-dual in the strict Aubry sense~\cite{AubryAndre1980}.  Instead, the real- and momentum-space equations belong to the same M\"obius--cosine nonlinear-eigenvalue class.  We call this relation \emph{class duality}.

The Liu--Xia condition can be solved exactly by treating the real- and dual-space transfer problems on equal footing.  In both representations the transfer matrix factorizes into a singular denominator and a regular analytic cocycle.  The two denominators,
\(E-t-2V\cos\theta\) in real space and \(E-2J\cos\theta\) in dual space, define two natural energy windows.  When
\begin{equation}
|E-t|\le 2|V|,
\qquad
|E|\le 2|J|,
\end{equation}
the singular denominator growth is exactly canceled by the normalized analytic cocycle.  Avila's complexified phase theory then places the normalized cocycles in the subcritical regime, where their Lyapunov exponents vanish on the real axis~\cite{Avila2,Avila1}.  The Liu--Xia condition therefore yields an exact finite critical region.

Outside this region, the system enters supercritical branches, where the criterion becomes a balance between real- and dual-space exponential growth rates.  Requiring both Lyapunov exponents to vanish selects the exact critical branch
\begin{equation}
E=\frac{Jt}{J-V}.
\end{equation}
Thus the class-dual self-dual line is elevated from a formal correspondence to an exact Liu--Xia critical branch.

Combining the subcritical region and the supercritical branch, the complete critical set is
\begin{equation}
\mathcal C
=
\mathcal C_{\mathrm{in}}
\cup
\mathcal C_{\mathrm{out}},
\label{eq:MC_complete_set_main}
\end{equation}
where
\(\mathcal C_{\mathrm{in}}=\{(E,V): |E-t|\le 2|V|,\ |E|\le 2|J|\}\) and
\(\mathcal C_{\mathrm{out}}=\{(E,V): E=Jt/(J-V),\ |E-t|>2|V|,\ |E|>2|J|\}\).
Figure~\ref{fig:dual_sector} displays this exactly predicted critical set in both spectra together with representative real- and dual-space wave-function profiles \(\psi_n\) and \(\phi_m\).  The transfer-matrix factorization, Avila complexification, and growth-rate matching are derived rigorously in Sec.~IV of the \SM.

\paragraph{Non-dual complex quasiperiodic potential model.---}
\begin{figure*}[t]
\centering
\includegraphics[width=0.98\textwidth]{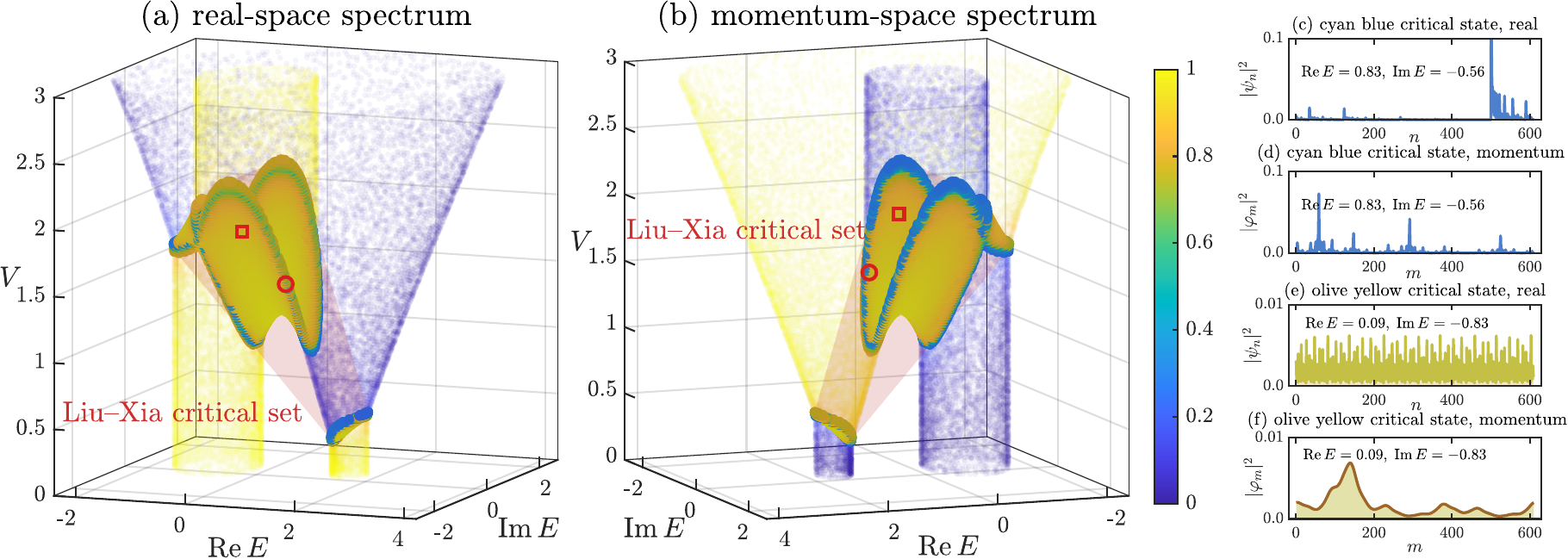}
\caption{
Liu--Xia critical set in a non-Hermitian complex-energy spectrum.
(a),(b) Real- and momentum-space spectra in \((\mathrm{Re}\,E,\mathrm{Im}\,E,V)\), colored by the corresponding fractal dimension \(D_2\).
The calculations use \(L=610\), \(\alpha=377/610\), \(J=t=g=1\), and \(\theta=0\).
The translucent red surface denotes the Liu--Xia critical set predicted by the exact conditions \(G_x(E)=G_m(E)=0\).
The red circle and square mark two representative critical states selected near \((E,V)=(1-0.5i,1.5)\) and \((0.1-1.0i,1.9)\), respectively, corresponding to small \(D_2\) (cyan-blue) and large \(D_2\) (olive-yellow). Panels (c)--(f) show their real- and momentum-space probability profiles.
}
\label{fig:nondual_complex}
\end{figure*}

Finally, we consider a minimal non-Hermitian quasiperiodic model with a complex one-sided modulation~\cite{Longhi1,Zhang,HatanoNelson1996,Yao2018},
\begin{equation}
\begin{aligned}
H
=
\sum_n
\Big[
&J a_{n+1}^{\dagger}a_n
+
g(a_n^{\dagger}b_n+b_n^{\dagger}a_n)
+
t b_n^{\dagger}b_n
\\
&+
V \e^{\ii(2\pi\alpha n+\theta)}
b_n^{\dagger}b_n
\Big],
\end{aligned}
\label{eq:nh_nondual_H_main}
\end{equation}
with \(J,V,g,t>0\).  The complex exponential potential and the unidirectional backbone hopping make the model non-Hermitian and remove any Aubry-type self-duality between real and momentum space.

Nevertheless, both representations remain exactly solvable at the Lyapunov level.  Eliminating \(b_n\) in real space gives
\begin{equation}
a_{n-1}
=
\frac{
E(E-t)-g^2
-
EV \e^{\ii(2\pi\alpha n+\theta)}
}{
J\left[
E-t-V \e^{\ii(2\pi\alpha n+\theta)}
\right]
}
a_n .
\label{eq:nh_nondual_real_recursion_main}
\end{equation}
In momentum space the complex exponential potential becomes a one-sided hopping, and eliminating \(a_k\) gives
\begin{equation}
b_{k-1}
=
\frac{
(E-t)
\left[
E-J \e^{\ii(2\pi\alpha k+\theta)}
\right]
-g^2
}{
V
\left[
E-J \e^{\ii(2\pi\alpha k+\theta)}
\right]
}
b_k .
\label{eq:nh_nondual_dual_recursion_main}
\end{equation}
Because both recursions are scalar, the Lyapunov drifts reduce to exact ergodic averages and can be evaluated by Jensen's formula~\cite{JensenFormula}.  For complex energy \(E\), the signed real- and momentum-space drifts are
\begin{equation}
\begin{aligned}
G_x(E)
&=
\ln
\frac{
\max\!\left\{
|E(E-t)-g^2|,
V|E|
\right\}
}{
J\,
\max\!\left\{
|E-t|,
V
\right\}
},
\\[6pt]
G_m(E)
&=
\ln
\frac{
\max\!\left\{
|E(E-t)-g^2|,
J|E-t|
\right\}
}{
V\,
\max\!\left\{
|E|,
J
\right\}
}.
\end{aligned}
\label{eq:nh_nondual_Gx_Gm_main}
\end{equation}
The nonnegative Lyapunov exponents are
\(\gamma_x(E)=|G_x(E)|\) and \(\gamma_m(E)=|G_m(E)|\).  The absolute value only removes the orientation dependence of the first-order recursion and does not affect the zeros.  Hence the Liu--Xia condition gives the critical set exactly as
\begin{equation}
\mathcal S_{\mathrm{crit}}
=
\left\{
(E,V)\in\mathbb C\times\mathbb R_+:
G_x(E)=0,
\quad
G_m(E)=0
\right\}.
\label{eq:nh_nondual_Scrit_main}
\end{equation}
Equivalently, the critical surface is fixed by the two closed-form equations
\begin{align}
\max\{|E(E-t)-g^2|,V|E|\}
&=J\max\{|E-t|,V\}, \\
\max\{|E(E-t)-g^2|,J|E-t|\}
&=V\max\{|E|,J\} .
\end{align}
This provides a genuinely non-self-dual test of the theory.  The critical surface is not obtained from an Aubry-type symmetry, but from the simultaneous vanishing of two independently solved Lyapunov exponents.  Figure~\ref{fig:nondual_complex} shows the exactly predicted complex-energy critical surface together with representative real- and Fourier-dual wave-function profiles \(\psi_n\) and \(\phi_m\).  The derivation is given in Sec.~V of the \SM.

\textit{Conclusion.---}
We have established localization criticality as a dual-space Lyapunov principle.  The key result is the Fourier exclusion principle: two Fourier-dual representations cannot both support finite exponential localization lengths.  The Liu--Xia condition,
\(\gamma_x(E)=\gamma_m(E)=0\),
therefore follows as a rigorous length-scale criterion rather than a conjectural diagnostic.  It identifies critical states by the simultaneous disappearance of exponential confinement in real and momentum space, while participation ratios and multifractal dimensions describe the remaining geometry within a chosen basis.

Beyond the solvable examples studied here, the significance of the Liu--Xia condition lies in its separation of criticality from basis-dependent wave-function morphology.  It suggests that the essential object is not a particular multifractal pattern in one representation, but the absence of exponential length scales in a pair of dual descriptions.  This viewpoint provides a natural route for extending critical-state diagnostics to systems where conventional single-particle notions of real and momentum space are no longer sufficient, including interacting quasiperiodic systems and many-body localization transitions~\cite{Basko2006,Imbrie2016,Nandkishore2015,Alet2018,Schreiber2015,Choi2016,Luschen2018}.  In such settings, the relevant dual spaces may be configuration space and its spectral or graph-dual representation, and a generalized Liu--Xia criterion may offer an invariant way to identify many-body critical states.

\end{document}